# AUTONOMIC HTML INTERFACE GENERATOR FOR WEB APPLICATIONS

## Youssef Bassil ,Mohammad Alwani

LACSC – Lebanese Association for Computational Sciences
Registered under No. 957, 2011, Beirut, Lebanon
youssef.bassil@lacsc.org , mohammad.alwani@lacsc.org

## ABSTRACT

*Recent advances in computing systems have led to a new digital era in which every area of life is nearly interrelated with information technology. However, with the trend towards large-scale IT systems, a new challenge has emerged. The complexity of IT systems is becoming an obstacle that hampers the manageability, operability, and maintainability of modern computing infrastructures. Autonomic computing popped up to provide an answer to these ever-growing pitfalls. Fundamentally, autonomic systems are self-configuring, self-healing, self-optimizing, and self-protecting; hence, they can automate all complex IT processes without human intervention. This paper proposes an autonomic HTML web-interface generator based on XML Schema and Style Sheet specifications for self-configuring graphical user interfaces of web applications. The goal of this autonomic generator is to automate the process of customizing GUI web-interfaces according to the ever-changing business rules, policies, and operating environment with the least IT labor involvement. The conducted experiments showed a successful automation of web interfaces customization that dynamically self-adapts to keep with the always-changing business requirements. Future research can improve upon the proposed solution so that it supports the self-configuring of not only web applications but also desktop applications.*

## 1. INTRODUCTION

Over the past decades, researches in computing science and information technology have been intensively carried out worldwide to pervade computer systems into every area of human life. In view of that, computers have revolutionized many of our daily work, processes, and activities, allowing both people and organizations to rely on programmed machines to automate complex and high-value tasks while manually handling trivial and routinely administrative and bureaucratic duties. However, the rapid changing demands of e-business and the ever-increasing competition in today's market place and global economy, accompanied by the myriad advances in computing technologies, have led to the rising of cutting-edge heterogeneous IT infrastructures and intricate high-tech environments. As a result, new drastic challenges came to light. In fact, the manageability, configurability, and maintainability of IT solutions are becoming more difficult to attain as the complexity of information systems and enterprise distributed infrastructures is escalating [1]. IBM stated that the complexity of emerging IT systems is an obstacle that prevents the industry from moving to the next era of computing and thereby is regarded as the most important upcoming challenge facing the realm of information technology [2]. This complexity is manifested in every aspect of IT processes, including hardware, software, network, programming languages, configuration, customization, and maintenance. For instance, IT administrators are constantly fighting to manage large and diverse computing architectures comprising hybrid components and spreading beyond a single organization. Additionally, they are struggling to setup and configure IT systems ran by users and employees belonging to different levels, different positions, and different departments. In effect, today's current corporate systems are no longer being deployed on local machines; instead, they are distributed over several servers, allowing users from all over the globe to connect to the same end-to-end service simultaneously. Further more, IT technicians are putting relentless effort to maintain and ensure the smooth operation of information systems because any system stoppage or breakdown can be severe on the running of the business and can cause direct financial losses. All in all, the complexity of nowadays IT systems is triggering off money losses, deficiencies in productivity, and delays in the implementation, operation, and troubleshooting of IT infrastructures.

The paramount challenge is to simplify IT processes and make them autonomous enough to answer dynamic customers' needs and the non-stop changing business demands. Autonomic computing systems can cope with this challenge as they are self-configuring, self-healing, self-optimizing, and self-protecting [1].

This paper proposes an autonomic web-interface generator for self-configuring Graphical User Interface (GUI) of web applications. The purpose of this system is to automate the customization of web interfaces in accordance with the business rules, policies, and operating environment with the minimum IT labor intervention. Since web applications are usually used by different classes and groups of users each with dissimilar authorization rights, access privileges, roles, and limitations, it is certainly a necessity to have several interface versions for the same application. Since it is impossible for IT people to keep pace with this degree of mixt constraints and requirements, it is a must to find a way to self-configure GUI interfaces of web applications with the least amount of manual work.

The proposed system is an autonomic HTML web-interface generator based on W3C XML Schema and a proprietary Style Sheet. Inherently, the process consists of parsing an XML Schema and a Style Sheet, and then translating them into an HTML web document containing web UI controls such as textboxes, dropdown lists, checkboxes, and radio buttons.The purpose of the Style Sheet is to format and decorate the generated output web-interface.

The cornerstone of the proposed autonomous generator is an arrangement of three components:a lexical analyzer also known as scanner, based on finite-state machine, responsible for recognizing sequence of characters and producing a sequence of meaningful tokens; a parser, based on context-free grammars (CFG),responsible for syntactically recognizing the XML Schema language; and a code generator based on attribute grammars (AG), responsible for generating the output HTML web-interface. The proposed solution is based on the de facto standards of formal language theory and XML technologies, and thus it can be generally integrated into any web solution to automate the customization of its user interfaces.

## 2. BACKGROUND

This section defines several key concepts and standards related to the work presented in this paper. They include the autonomic system and its properties, the XML language, and the XML Schema.

## 2.1. Autonomic System

An autonomic computing system is a proactive system that automatically senses, analyzes, adjusts, and adapts itself based on the business requirements, policies, and computing environment, relieving IT professionals from performing laborious and manual tasks [3].Fundamentally, an autonomic system has four characteristics: self-configuring, self-healing, self-optimizing, and self-protecting [1, 3]. Figure 1 depicts these four characteristics.

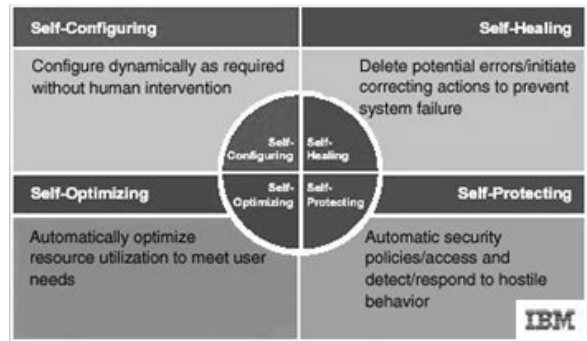

Figure 1.Characteristics of an autonomic system (Courtesy of IBM)

Self-configuring: An autonomic system automatically changes its configuration according to the operating conditions of its computing environment. Some examples may include setup software automatically, downloading and installing software patches and service packs, self-customizing user interfaces, and updating virus signatures.

Self-healing: An autonomic system automatically discovers and corrects hazards and errors, and recovers from failures without human intervention. Some examples may include promoting loosely coupled systems, promoting self-restarting services and fault-tolerant modules, creating backup plans and replicas for existing components and using them for a failure scenario.

Self-optimizing: An autonomic system automatically tunes its initial setup and acts proactively to adapt itself with respect to the unexpected changes in its execution environment. Some examples may include increasing automatically disk storage to cope with data growth, applying load balancing techniques to contain the increase of users, assigning extra processing cycles to computationally intense applications, and requesting more Internet bandwidth for rich-Internet applications.

Self-protecting: An autonomic system automatically detects, analyzes, prevents, and defends against security threats, malwares, hostile attacks, and unauthorized access. Some examples may include encrypting communication lines between various nodes of the system, deploying intrusion detection and prevention systems, isolating unsolicited system behaviors, and forcing policies and user access controls.

## 2.2. XML and XML Schema

XML short for Extensible Markup Language is an open standard for exchanging data between computing systems [4]. Basically, XML is a plaintext portable language made out of elements and attributes that define the semantics of data. XML is case-sensitive and is understandable by both humans through a traditional text editor, and computers through an XML parser.

Style Sheets are also plaintext files that unlike XML language which is data-oriented, they are formatting-oriented in that they contain instructions to format and render XML files into a web browser. These instructions dictate the color of text, the font size, and other layout-related items for the XML document.

W3C XML Schema is an XML-based language recognized by the World Wide Web Consortium (W3C) community, used to describe the structure of an XML document [5]. Actually, an XML Schema dictates the elements, their attributes, their data types, and their number of occurrence in a given XML file. Any violation of these rules would make the corresponding XML file invalid and ill-formed. Following is an XML file adhering to the specifications of an XML Schema.

```
<!— XML Schema —>
<xs:schemaxmlns:xs="http://www.w3.org/2001/XMLSchema">
<xs:element name="Book">
<xs:complexType>
<xs:sequence>
<xs:element name="Title" type="xs:string" />
<xs:element name="Publisher" type="xs:string" />
<xs:element name="Price" type="xs:decimal" />
</xs:sequence>
</xs:complexType>
</xs:element>
</xs:schema>

<!— XML Document —>
<Book>
<Title>C# How to Program</ Title>
<Publisher>Prentice Hall</ Publisher>
<Price>45.17</Price>
</Book>
```

## 3. RELATED WORK

Various techniques and solutions have been developed to translate definition files such as XML Schema into web user interfaces and HTML documents. In that sense, Kuikka and Penttonen[6] proposed a transformation method based on document type definition (DTD)which converts document instances originally bounded to generic DTDs into another document types. In this approach an already existing document X along with its grammar Y and a new grammar Y'are fed to the system. The system then converts X into a corresponding X' that sticks to the new grammar Y'. That way, using a grammar Y' that defines HTML elements can easily transform a DTD into a web interface. Amaneddine, Bahsoun, and Bodeveix[7] proposed TransM, a structured document transformation model whose purpose is to convert an instance document conforming to a DTD into another instance document conforming to a another DTD derived from the transformation rules of the TransM system. Additionally, the system harnesses the XSLT transformation language to format and apply styles on the output document that eventually can be rendered as a web interface.Bird, Good child, and Halpin[8] proposed an approach to transform meta-models into XMLSchemas and vice versa with minimum data redundancy. The input model is versatile in that it describes any type of systems including web interfaces. For this purpose, an algorithm based on twelve heuristic rules is used to assist the conversion of metadata into XML Schema. Psaila and Grespi-Reghizzi[9]proposeda document level strategy to directly associate semantic definitions with DTD files. In this approach, a new specification language for DTD semantics is derived called SRD (Semantics Rules Definitions) which associate severy element in a DTD with a semantic feature. Then, a semantic evaluator transforms the SRD document into an XML generator able to produce a decorated XML file. Ramalho, Lopes, and Henriques[10]proposed a context-free grammar for DTD documents. The technique pivots around formally transforming DTD documents into an attribute grammar [11] that can produce an SGML editor capable of generating HTML interfaces. Hanna and Frost [12] introduced a translator that converts DTDs into Voice XML applications. Voice XML is a markup language that drives and controls applications based on voice recognition. The translator is at the core based on context-free grammars that manage the conversion process. Lift [13] is a web programming framework that employs a linking engine that uses templates to transform scripting instructions into web pages. Lift also offers advanced utilities and functions to automatically customize the output being generated. Pohjalainen[14] proposed a method for self-configuration of user interface via software introspection combined with semantic mapping of backend methods that eases the management of user-interface during the development process.Penner[15] proposed a tool that integrates various subsystems and automatically generates and configures graphical user interfaces. The approach employs object-oriented models of interface elements and interactions, managed by an algorithm that dynamically generates user interfaces pertaining to particular users. Scholaert[16] proposed an autonomous man-machine interface for a communication device in a network environment that can be dynamically updated according to predefined computing conditions. The man-machine interface is constantly updated to reflect services offered by service providers in a specific operational environment.

## 4. AUTONOMIC WEB-INTERFACE GENERATOR

This paper proposes an autonomic web-interface generator that converts declarations in an XML Schema and styles in a Style Sheet into a styled HTML web page, enclosing user-interface controls such as textboxes and buttons. Additionally, the system can generate JavaScript check routines that validate the correctness of data that are input to the generated web-interface. At heart, the proposed autonomic system is a compiler system that consists of a lexical analyzer, a parser, and a code generator. The goal of the proposed design is to automate the customization of web interfaces; thus freeing IT administrators from the burden ofre-designing web user-interfaces whenever the business policies, system requirements, and operating environment change. The proposed autonomic system comprises several stages of operation :The first stage is feeding the system with the input files, mainly the XML Schema file enclosing web-interface specifications and the Style Sheet file enclosing styles to format the output web-interface. The second stage is scanning the sequence of characters from both input files and converting them into meaningful tokens.The third stage is parsing the previously produced list of tokens so as to determine whether or not they conform to the grammar of the system; the fourth stage is converting the previously parsed sequence of tokens into a styled web-interface, mainly represented as an HTML web page.Figure 2 shows the different stages of the proposed autonomic web-interface generator.

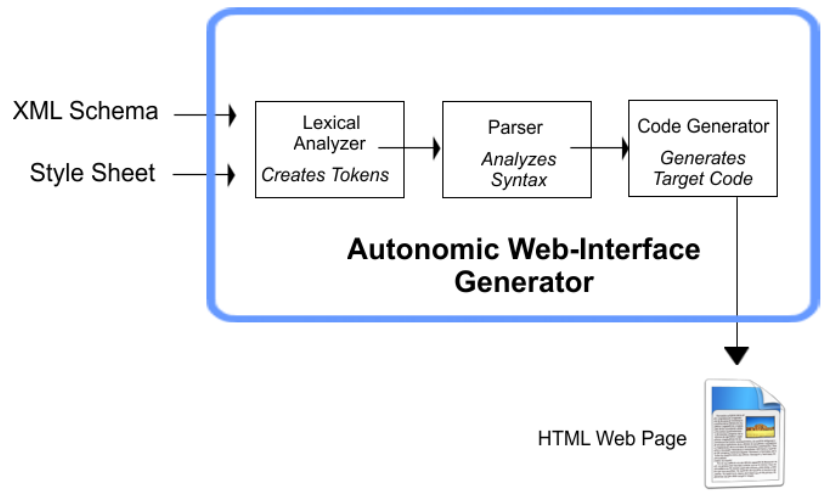

Figure 2.Different stages of the proposed autonomic web-interface generator

## 4.1. XML Schema Specifications

By definition, an XML Schema specifies what elements must and must not occur in an XML file. The XML Schema employed in the proposed solution is originally derived from the draft of W3C XML Schema standard [17], however modified in order to meet the requirements of the proposed generator, and it is defined by the following rules:

1. Elements can be of simple or complex type, and they both have a name and a data type.
2. Simple-type elements can only contain alphanumeric values having valid data types, and they cannot contain nested child elements or attributes.
3. Complex-type elements can contain nested child elements and attributes.
4. Elements must appear only once, however the *minOccurs* and *maxOccurs* attributes (default value 1) can be used to control the number of occurrence of a given element.
5. Data types are all reserved primitive types and they are as follows:

   a) string
   b) integer
   c) positiveInteger
   d) boolean
   e) decimal
   f) dateTime

6. New data types can be derived by applying restrictions on simple data types. These restrictions are created with the help of facets, and they are listed below:

   a) length
   b) minLength
   c) maxLength
   d) pattern
   e) enumeration
   f) totalDigits
   g) fractionDigits

## 4.2. Style Sheet Specifications

The Style Sheet employed in the proposed system is a proprietary style sheet loosely based on Cascading Style Sheets (CSS). Its purpose is to specify the type of UI controls to be generated and their formatting styles.

Essentially, the Style Sheet has only one basic instruction that contains several variables, as well as several optional fields. Below is its syntax and specifications:

**element:***element-name*(**tag:***tag-name* , [**listype:***list-type*]**)**

The *element-name* specifies the element in the XML schema that needs to be converted into an HTML user control. The *tag-name* specifies the actual HTML user control that the *element-name* is to be converted to. The *tag-name* is responsible for formatting the *element-name* because it is graphically rendered by the web browser. The *list-type* is optional and is only valid when the *tag-name* is equal to *ol* (ordered list), and specifies the type of ordered numbering (e.g. 1,2,3 - a,b,c - i,ii,iii).The instructions of the style sheet are executed sequentially and their number must match the number of elements in the XML schema. Any element that is not required to be rendered must have a *tag-name* equals to *null*. Below is the list of the various *tag-name* that can be recognized by the system:

Headers:

StyleSheet Syntax:

    **element:***element-name* (**tag:**h1)
    **element:***element-name* (**tag:**h2)
    **element:***element-name* (**tag:**h3)
    **element:***element-name* (**tag:**h4)
    **element:***element-name* (**tag:**h5)
    **element:***element-name* (**tag:**h6)

Equivalent rendered HTML:

    <h1>*value*</h1>
    <h2>*value*</h2>
    <h3>*value*</h3>
    <h4>*value*</h4>
    <h5>*value*</h5>
    <h6>*value*</h6>

The *value* of a header is actually extracted from the *default* attribute of the corresponding XML Schema element.

Ordered lists and Unordered lists:

Style Sheet syntax:

    **element:***element-name* (**tag:**ol , **listype:***list-type*)
    **element:***element-name* (**tag:**ul)

Equivalent rendered HTML:

```
<ol type ="list-type">
<li>value</li>
<li>value</li>
</ol>

<ul>
<li>value</li>
<li>value</li>
</ul>
```

Only complex-type elements can be rendered as ordered and unordered lists. Every nested child element is rendered as list item <li>. The variable *list-type* specifies the type of ordered numbering (e.g. 1,2,3 - a,b,c - i,ii,iii). The *value* of a list item is actually extracted from the *default* attribute of the corresponding XML Schema child element.

Anchor:

Style Sheet syntax: **element**:*element-name* (**tag**:a)

Equivalent rendered HTML: <a href="*h-value*">*value*</a>

The *h-value* is actually extracted from the *fixed* attribute of the corresponding XML Schema element, whereas *value* is extracted from the *default* attribute of the same element.

Label:

Style Sheet syntax: **element**:*element-name* (**tag**:label)

Equivalent rendered HTML: <label>*value*</label>

The *value* of a label is actually extracted from the *default* attribute of the corresponding XML Schema element. This type of element is usually rendered as a plain text.

Input Controls:

Style Sheet syntax:

```
element:element-name (tag:text)
element:element-name (tag:password)
element:element-name (tag:checkbox)
element:element-name (tag:radio)
element:element-name (tag:submit)
element:element-name (tag:textarea)
element:element-name (tag:select)
```

Equivalent rendered HTML:

```
<input name="element-name" type="text" value="value" />
<input name="element-name" type="password" value="value" />
<input name="element-name" type="checkbox" value="value" />
<input name="element-name" type="radio" value="value" />
<input name="element-name" type="submit" value="value" />
<textarea name="element-name">value</textarea>
```

```
<select name="element-name">
<option>value</option>
<option>value</option>
</select >
```

The *value* of the input control is actually extracted from the *default* attribute of the corresponding XML Schema element. Since input controls allow users to input arbitrary data, JavaScript check routines are generated automatically as part of the output HTML web-interface, to validate the data entries against the data types and facets specified in the XML Schema. The data types are: *string*, *integer*, *positive Integer*, *boolean*, *decimal*, *dateTime*, and the facets are: *length*, *minLength*, *maxLength*, *pattern*, *enumeration*, *totalDigits*, *fractionDigits*.

Only elements with *restriction* and *enumeration* facets can be rendered as *select* control. Every enumeration is actually rendered as a select item <option>. The *value* of every select item is actually extracted from the values of the corresponding enumeration elements.

Null Element:

Style Sheet syntax: **element**:*element-name* (**tag**:null)

This indicates that the associated *element-name*should not be rendered in the web-interface.

## 4.3The Lexical Analyzer

The purpose of the lexical analyzer is to scan the stream of characters received from the input XML Schema and the Style Sheet, and produce another stream of meaningful units called tokens [18]. Tokens can be divided into two categories: keywords and symbols.
Keywords for the XML Schema are:

*schema, xs, element, name, type, fixed, default, complexType, simpleType, sequence, minOccurs, maxOccurs, base, restriction, value, string, integer, positiveInteger, boolean, decimal, dateTime, length, minLength, maxLength, pattern, enumeration, totalDigits, fractionDigits.*

Symbols for the XML Schema are:

<>: = " / + -

Keywords for the Style Sheet are:

*element, tag, h1, h2, h3, h4, h5, h6, ol, type, ul, a, label, text, password, checkbox, radio, textarea, select, null.*

Symbols for the Style Sheet are:

: ( )

Other tokens are *identifier*, *numerical-value*, *string-value*, *letter*, and *digit*, and are defined by the following regular expressions:

identifier = letter (digit | letter)*
numerical-value = (+ | -)? digit digit* (. digit digit*)?
string-value = (letter | digit)*
letter = a|..|z|A|..|Z
digit = 0|..|9

Uppercase and lowercase letters are distinct and therefore *complexType* is different from *complextype* and *AuthorName* is different from *authorName*.

Whitespaces are discarded by the lexical analyzer and are usually represented by blank spaces, carriage returns, new lines, and tabs.Moreover, whitespaces cannot occur within a token word.

## 4.4The Parser and Its Context-Free Grammar

The purpose of the parser is to determine whether or not the source tokens are arranged according to the rules of the context-free grammar of the language. The context-free grammar (CFG)originally proposed by [19], describes the structure and syntax of the XML Schema and the Style Sheet languages, and can be stated as G=(V, T, R, S) where V is the set of non-terminals, T is the set of terminals, R is the set of production rules, and S is the start non-terminal variable. The production rules of the CFG for parsing the XML Schema are as follows:

Schema→**<xs:schema>**Element-List**</xs:schema>**
Element-List→Simple-Type | Complex-Type
Simple-Type →**<xs:element** Element-Attributes**/>**Element-List
Complex-Type →**<xs:element**Element-Name**>**
**<xs:complexType**Element-Name>**<xs:sequence** Sequence-Attributes >
Element-List | DerivedDataType-Element
**</xs:sequence></xs:complexType>**
**</xs:element>**Element-List
Element-List →ε
DerivedDataType-Element→**<xs:element** Element-Name>|ε
**<xs:simpleType** Element-Name >
**<xs:restriction base="PrimitiveDataType">**
Facets-List
**</xs:restriction>**
**</xs:simpleType>**
**</xs:element>**DerivedDataType-Element
Facets-List→**<xs:Facetsvalue="string-value"/>**|Facets-List
Facets-List →ε
Sequence-Attributes →**minOccurs = "string-value" maxOccurs = "string-value"** |ε
Element-Attributes→ Element-Name Element-DataType Element-Optional-Attributes|ε
Element-Optional-Attributes→**default="string-value"**| **fixed="string-value"**|ε
Element-Name →**name="identifier"**
Element-DataType→ **type="PrimitiveDataType"**
PrimitiveDataType→**xs:string | xs:integer | xs:positiveInteger | xs:boolean | xs:decimal| xs:dateTime**
Facets→**length | minLength | maxLength | pattern | enumeration | totalDigits | fractionDigits**

The production rules of the CFG for parsing the Style Sheet are as follows:

StyleSheet→Element-List
Element-List →**element:identifier** (**tag**:Tag-Name) Element-List
Tag-Name→**h1| h2 | h3 | h4 | h5 | h6 | ol,type**:Type **| ul | a | label | text | password | checkbox | radio
| textarea | select | null**
Type →**"l" | "1" | "a" | "i"**
Element-List →ε

## 5. EXPERIMENTS AND RESULTS

In the experiments, the proposed autonomic web-interface generator was evaluated. The test data were an XML Schema and a Style Sheet for specifying a web-interface for creating generic user accounts. The autonomic system successfully generated the appropriate HTML web page from the input files. Furthermore, the user controls specifications in the same XML Schema we realtered, and the input files were fed again to the generator which effectively managed to re-generate the corresponding web page with different user controls complying with the new input declarations. Below is the original XML Schema containing specifications of a web-interface for creating user accounts.

```
<xs:schema>
<xs:element name="title" type="xs:string" default="Fill-in the below fields"/>
<xs:element name="fullName1" type="xs:string"default="Enter your Full Name: " />
<xs:element name="fullName2" type="xs:string"/>
<xs:element name="dateofBirth2" type="xs:string"default="Enter your Date of Birth: " />
<xs:element name="dateofBirth2" type="xs:dateTime" default="1/1/2010"/>
<xs:element name="pass1" type="xs:string"default="Enter your Password: " />
<xs:element name="pass2" type="xs:string"/>
<xs:element name="phone1" type="xs:string"default="Enter your Phone Number: " />
<xs:element>
<xs:complexType>
<xs:sequence>
<xs:element name="phone2">
        <xs:simpleType>
          <xs:restriction base="xs:string">
          <xs:minLength value="6"/>
          <xs:maxLength value="12"/>
          </xs:restriction>
        </xs:simpleType>
  </xs:element>
<xs:element name="country1" type="xs:string"default="Select your Country: " />
 <xs:element name="country2">
 <xs:simpleType>
 <xs:restriction base="xs:string">
 <xs:enumeration value="United States" />
 <xs:enumeration value="France" />
 <xs:enumeration value="UK" />
 <xs:enumeration value="Canada" />
 </xs:restriction>
 </xs:simpleType>
 </xs:element>
</xs:sequence>
</xs:complexType>
</xs:element>
<xs:element name="hiddenField" type="xs:string" />
<xs:element name="comments1" type="xs:string"default="Enter Your Comments: " />
<xs:element name="comments2" type="xs:string" default="Your comments go here"/>
<xs:element name="support" type="xs:string" default="Contact our Support" fixed="support.html"/>
<xs:element name="submit" type="xs:string"default="Click to Submit" />
</xs:schema>
```

Next, is the Style Sheet containing specifications for formatting the above XML Schema. The Element *title* is to be converted to a header *<h2>*;elements *title*, *fullname1*, *dateofBirth1*, *pass1*, *phone1*, *country1*, and *comments1* are to be converted to *<label>*;elements *fullname2*, *dateofBirth2*, and *phone2* are to be converted to textbox controls*<input type="text"/>*;element *pass2* is to be converted to a password control *<input type="password"/>*; element *country2*is to be converted to a select control *<select>*; element *comments2* is to be converted to a textarea control *<textarea>*;element *support* is to be converted to an anchor *<a>*; and element *submit* is to be converted to a submit button control *<input type="submit"/>*. In addition of generating GUI elements, a JavaScript check routine is to be generated to handle the derived data type of element *phone2*.

**element**:title (**tag**:h2)
**element**:fullname1 (**tag**:label)
**element**:fullname2 (**tag**:text)
**element**:dateofBirth1 (**tag**:label)
**element**:dateofBirth2 (**tag**:text)
**element**:pass1 (**tag**:label)
**element**:pass2 (**tag**:password)
**element**:phone1 (**tag**:label)
**element**:phone2 (**tag**:text)
**element**:country1 (**tag**:label)
**element**:country2 (**tag**:select)
**element**:hiddenField (**tag**:null)
**element**:comments1 (**tag**:label)
**element**:comments2 (**tag**:textarea)
**element**:support (**tag**:a)
**element**:submit (**tag**:submit)

The following HTML code represents the source-code of the generated web page. It contains all the elements that were stipulated in the XML Schema and formatted by the Style Sheet, in addition to a JavaScript check routine that validates the input length of the element *phone*. Figure 3depicts the actual generated HTML web page loaded in a web browser.

```
<html>
<head>
<script type="text/javascript">
functionvalidate_phone(object_var)
    {
if(object_var.value.length< 6 || object_var.value.length> 12)
object_var.value = "Bad Input" ;
    }
</script>
</head>
<body>
<form>
<h2>Fill-in the below fields</h2>
    Enter your Full Name: <input type="text" name="fullname2"/>
<br>
    Enter your Date of Birth: <input type="text" name="dateofBirth2" value="1/1/2010"/>
<br>
    Enter your Password: <input type="password" name="pass2"/>
<br>
    Enter your Phone Number: <input type="text" name="phone2" onBlur="validate_phone(this)"/>
<br>
    Select your Country: <select name="country2">
<option>United States</option>
<option>France</option>
<option>UK</option>
<option>Canada</option>
</select>
<br>
    Enter Your Comments: <textarea name="comments2" rows="5" cols="30">
your comments go here</textarea>
<br>
<a href="support.html">Contact our Support</a>
```

```
<br>
<input type="submit" name="submit" value="Click to Submit"/>
</form>
</body>
</html>
```

Figure 3.The generated HTML web page

In order to test the self-configuring property of the proposed generator, the XML Schema and the Style Sheet were modified: elements *dateofBirth1, dateofBirth2, phone1, phone2,* and *support*were removed; the header size of element *title* was changed from *<h2>* to *<h4>;*two new elements were introduced, one of type *label* called *email1* and one of type *text* called *email2*;the caption of the submit element was altered; and the *comments2* element was transformed from a *textarea* control into a *text* control. The new results obtained are depicted in Figure 4.

Figure 4.The re-generated HTML web page

## 6. CONCLUSIONS AND FUTURE WORK

This paper presented an autonomic HTML web-interface generator that transforms declarations in XML Schema into HTML web pages encompassing GUI elements and user controls. Besides, a proprietary Style Sheet was used to render and format the visual presentation of the output web-interface. The system is based on a compiler majorly composed of a lexical analyzer, a CFG parser, and a code generator. Altogether, they excelled in automating the process of generating web interfaces from XML Schema and Style Sheet files. As a result, IT people can now specify the requirements for their systems' interfaces through writing an XML Schema file and formatting it through a Style Sheet. Modifying the application's interface would only require editing the XML Schema. No re-compilation or re-deployment of the original application is to take place. Consequently, IT people are dismissed from the load of maintaining, managing, and adapting GUI web interfaces according to the ever-changing computing environment and business requirements. Likewise, the overall development and maintenance costs and time are to be reduced which will simplify the implementation of rapidly growing computing infrastructures. As future work, the language of the proposed generator is to be extended so as to deliver richer multimedia-based web-interfaces. Moreover, self-configuring of graphical interfaces for desktop and mobile applications is to be investigated so that to support different types of IT infrastructures running on different computing platforms.

## ACKNOWLEDGMENTS

This research was funded by the Lebanese Association for Computational Sciences (LACSC), Beirut, Lebanon under the "Autonomic Computing Research Project – ACRP2011".

## REFERENCES

[1]   Richard Murch, (2004)Autonomic Computing, Prentice Hall.

[2]   Paul Horn, (2001)Autonomic Computing, IBM's Perspective on the State of Information Technology, IBM Corporation.

[3]   Parashar,M. &Hariri,S., (2007)Autonomic computing: concepts, infrastructure, and applications, CRC Press.

[4]   Erik T. Ray, (2003)Learning XML, 2nded, O'Reilly Media.

[5]   Deitel, P., Deitel, H., Deitel, A.(2011)Internet & World Wide Web: How to Program, 5thed, Prentice Hall.

[6]   Kuikka, E. &Penttonen, M., (1993)"Transformation of Structured Documents with the Use of Grammar", Electronic Publishing, Vol. 6, No.4, pp373-383.

[7]   Amaneddine, N., Bahsoun, J.P. &Bodeveix, J.P., (2004) "TransM: A Structured Document Transformation Model", Proceeding of the 3rd International Conference Information Systems Technology and its Applications (ISTA), Salt Lake City, Utah.

[8]   Bird, L.,Goodchild, A.,&Halpin, T.,(2000)"Object Role Modelling and XML-Schema", In International Conference on Conceptual Modelling (ER), Salt Lake City, UT.

[9]   Psaila, G. &Grespi-Reghizzi, S., (1999) "Adding semantics to XML", Proceeding of Second Workshop on Attribute Grammars and their Applications, WAGA99, pp113-132, INIRIA, Paris.

[10]  Ramalho, J.C., Lopes, A. R. &Henriques, P., (1998) "Generating SGML specific Editors: from DTDs to Attribute Grammars", Proceeding of the Markup Technologies '98 Conference, pp61-72, Chicago, USA.

[11]  Reps, W., &Teitelbaum, T., (1989) "The Synthesizer Generator: A System for Constructing Language-Based Editors", Spinger-Verlag,.

[12]  Hanna, F. & Frost, R., (2007) "Adding Semantics to Formal Data Specifications to Automatically Generate Corresponding Voice Data-input Applications", Proceeding of the IEEE First International Conference on Semantic Computing (ICSC), pp405-412, California, USA.

[13]  Chen-Becker, D.,Weir, T., &Danciu, M., (2009)The Definitive Guide to Lift: A Scala-based Web Framework, Apress, Berkely, CA, USA.

[14]  Pohjalainen, P., (2010) "Self-configuring User Interface Components", Proceedings of the First Workshop on Semantic Models for Adaptive Interactive Systems (SEMAIS).

[15] Penner, R., (2009)"Self-Configuring User Interface Design", Sol No.: Navy SBIR FY2009.1, Firm: Iterativity, Inc., 3236 17th Ave. S. Minneapolis, Minnesota 55407.

[16] Scholaert, H., (2010) "Self-Configuring Man-Machine Interface for a Communication Terminal", United States Patent No: 20100248719.

[17] W3C XML Schema draft standard, (2011), latest revision, http://www.w3.org/XML/Schema

[18] Cooper, K. &Torcson, L. (2011) "Engineering a Compiler", 2nd ed., Morgan Kaufmann publishers.

[19] Chomsky Noam, (1956) "Three models for the description of language", Information Theory, IEEE TransactionsVol. 2, No. 3, pp113–124.